\documentclass[11pt,twoside,a4paper]{article}
\catcode`\@=11
\usepackage{rawfonts,amsmath,amsfonts,amssymb,epsfig,color}
\newtheorem{theo}{Theorem}

\newtheorem{lem}{Lemma}
\def\QED{\hbox{\hskip 1pt \vrule width4pt height 6pt depth 1.5pt \hskip 1pt}}

\title{On $LR(k)$-parsers of polynomial size}

\author{Norbert Blum
\thanks{Institut f\"ur Informatik, Universit\"at Bonn,
        Friedrich-Ebert-Allee 144, 
        D-53113 Bonn, Germany, 
        email:{ \tt blum@cs.uni-bonn.de}.
{An extended abstract of this paper appeared in {\em 37th International Colloquium,
ICALP 2010, Bordeaux, France, July 2010, Proceedings, Part II, LNCS 6199, pp. 163--174,
Springer-Verlag Berlin Heidelberg 2010.}}} }

\begin{document}
\date{October 15, 2013}
\maketitle


\begin{abstract}
Usually, a parser for an $LR(k)$-grammar $G$ is a deterministic pushdown 
transducer which produces backwards the unique rightmost derivation for a given input 
string $x \in L(G)$. The best known upper bound for the size of such a parser is
$O(2^{|G||\Sigma|^k+k\log |\Sigma| + \log |G|})$ where $|G|$ and $|\Sigma|$ are the sizes of the
grammar $G$ and the terminal alphabet $\Sigma$, respectively. If we add to a parser the 
possibility to
manipulate a directed graph of size $O(|G|n)$ where $n$ is the length of the input then
we obtain an extended parser. The graph is used for an efficient parallel simulation of all 
potential leftmost derivations of the current right sentential form such that the unique
rightmost derivation of the input can be computed. Given an arbitrary $LR(k)$-grammar 
$G$, we show how to construct an extended parser of $O(|G| + \#LA |N|2^k k \log k)$ size where
$|N|$ is the number of nonterminal symbols and $\#LA$ is the number of relevant
lookaheads with respect to the grammar $G$.
As the usual parser, this extended parser uses
only tables as data structure. Using some ingenious data structures and increasing the parsing
time by a small constant factor, the size of the extended
parser can be reduced to $O(|G| + \#LA|N|k^2)$. The parsing time is $O(ld(input) + k|G|n)$ 
where $ld(input)$ is the length of the derivation of the input. Moreover,
we have constructed a one pass parser. 
\end{abstract}

 \section{Introduction}
 Efficient implementations of parsers for context-free grammars play an 
 important role with respect to the construction of compilers. Since practical
 algorithms for general context-free analysis need cubic time, during the
 sixties subclasses of the context-free grammars having linear time parsers 
 have been defined. The most important such subclasses are the $LR(k)$-
 and the $LL(k)$-grammars. But the size of linear $LR(k)$- and $LL(k)$-parsers
 might be exponential in the size of the underlying grammar. Indeed, Ukkonen
 \cite{Uk} has constructed families of $LR(0)$- and $LL(2)$-grammars having
 only parsers of exponential size. The reason is that parsers read the input from left
 to right in one pass without backtrack and treat always the only possible 
 derivation which can depend on the prefix of the input derived so far. Hence, the
 state of the parser has to include all necessary information about the prefix of the input
read so far. Instead of the 
 treatment of the only possible derivation one can consider a set of potential 
 derivations which always contains the correct derivation in parallel. Hence, the
 following question arises: Is it possible to simulate an accurate set of derivations
 in parallel such that the correct derivation will be computed, the needed time remains
 linear and the modified parser uses on the input one pass without backtrack and has
 only polynomial size?

 In \cite{Bl2} for $LL(k)$-grammars the following positive answer to this question is
 given: If we add to a parser the possibility to manipulate a constant number of 
 pointers which point to positions within the constructed part of the leftmost 
 derivation and to change the output in such positions, we obtain an extended parser
 for an $LL(k)$-grammar $G$. Given an arbitrary $LL(k)$-grammar $G = (V, \Sigma,P,S)$,
 it is shown how to construct an extended parser of size $O(|G| + k|N||\Sigma|^k)$ 
 manipulating at most $k^2$ pointers. The
 parsing time is bounded by $O(n)$ where $n$ is the length of the input.
 In the case of $LR(k)$-grammers the situation is a little bit more complicated. The 
 parser has to take into account all possible derivations of the current right 
 sentential form. Hence, the state of the parser has to include all necessary 
 information with respect to all possible derivations of the current right
 sentential form from the start symbol. Instead of storing the whole needed 
 information into the state the parser can treat simultaneously all potential leftmost
 derivations and also backwards the rightmost derivation which has to be computed.
 Hence, with respect to $LR(k)$-grammars, the following question arises: For the 
 computation of the unique rightmost derivation, is it possible to simulate all 
 possible leftmost derivations and also backwards the rightmost derivation such that 
 the rightmost derivation will be computed, the needed time remains linear, the 
 modified parser uses on the input one pass without backtrack and has only polynomial 
 size?

 We will consider arbitrary $LR(k)$-grammars. The usual parser for an $LR(k)$-grammar 
 $G = (V,\Sigma ,P,S)$ is the so-called {\em canonical $LR(k)$-parser\/}. The best known 
 upper bound for its size is $O(2^{|G||\Sigma|^k + k\log|\Sigma| + \log |G|})$ \cite{SiSo2}. Hence, 
 DeRemer \cite{DeRe} has 
 defined two subclasses of the class of $LR(k)$-grammars, the $SLR(k)$-grammars and 
 the $LALR(k)$-grammars. Both classes allow smaller canonical $LR$-parsers. But the 
 size of these parsers can still be $O(2^{|G|})$ \cite{SiSo2}. Hence, the question posed above 
 remains interesting for $SLR(k)$- and $LALR(k)$-grammers, too. We will give a positive 
 answer to this question for arbitrary $LR(k)$-grammars.
 We assume that the reader is familiar with the elementary theory of $LR(k)$-parsing 
 as written in standard text books (see e.g. \cite{AhUl,Bl1,Ch,GrJa,SiSo1,WiMa}).
 First, we will review the notations used in the subsequence.

\section{Basic Notations}

 A {\em context-free grammar (cfg)\/} $G$ is a four-tuple $(V,\Sigma,P,S)$ where
 $V$ is a finite, nonempty set of symbols called the {\em total vocabulary\/},
 $\Sigma \subset V$ is a finite set of {\em terminal symbols\/}, 
 $N := V \setminus \Sigma$ is the set of {\em nonterminal symbols\/} 
 (or {\em variables\/}), $P$ is a finite set of  {\em productions\/}, and $S \in N$ is 
 the {\em start symbol\/}. The productions are of the form 
 $A \rightarrow \alpha$, where $A \in N$ and $\alpha \in V^*$. 
 $\alpha$ is called an {\em alternative\/} of $A$. $L(G)$ denotes 
 the {\em context-free language\/} generated by $G$.
 The {\em size\/} $|G|$ of the cfg $G$ is defined by
 $|G| := \sum_{A \rightarrow \alpha \in P} lg(A \alpha),$
 where $lg(A \alpha)$ is the length of the string $A \alpha$.
 As usual, $\varepsilon$ denotes the empty word. 
 A derivation is {\em rightmost\/} if at each step a production is applied to the 
 rightmost variable. A sentential form within a rightmost derivation starting in $S$ is 
 called {\em right sentential form\/}. {\em Leftmost\/} derivation and {\em left 
 sentential form\/} are defined analogously.
 A context-free grammar $G$ is {\em ambiguous\/} if there exists $x \in L(G)$ 
 such that there are two distinct leftmost derivations of $x$ from the start 
 symbol $S$. A context-free grammar $G = (V, \Sigma, P, S)$ is {\em reduced\/} 
 if $P = \emptyset$ or, for each $A \in V$, $S \stackrel{*}{\Rightarrow} \alpha A\beta 
 \stackrel{*}{\Rightarrow} w$ for some $\alpha ,\beta \in V^*,w \in \Sigma^*$.
In the subsequence, all derivations will be rightmost.

 A {\em pushdown automaton\/} $M$ is a seven-tuple $M = (Q,\Sigma,
 \Gamma,\delta,$ $q_0,Z_0,F)$, where $Q$ is a finite, nonempty set of {\em 
 states}, $\Sigma$ is a finite, nonempty set of {\em input symbols}, $\Gamma$
 is a finite, nonempty set of {\em pushdown symbols}, $q_0 \in Q$ is the
 {\em initial state}, $Z_0 \in \Gamma$ is the {\em start symbol\/} of the 
 pushdown store, $F \subseteq Q$ is the set of {\em final states}, and
 $\delta$ is a mapping from $Q \times (\Sigma \cup \{\varepsilon\}) \times
 \Gamma$ to finite subsets of $Q \times \Gamma^*$.
 A pushdown automaton is {\em deterministic\/} if for each $q \in Q$ and $Z \in 
 \Gamma$ either
 $\delta(q,a,Z)$ contains at most one element for each $a \in \Sigma$ and
 $\delta(q,\varepsilon,Z) = \emptyset$ or
 $\delta(q,a,Z) = \emptyset$ for all $a \in \Sigma$ and 
 $\delta(q,\varepsilon,Z)$ contains at most one element.
 A deterministic pushdown tranducer is a deterministic pushdown automaton 
 with the additional property to produce an output. More formally, a 
 {\em deterministic pushdown tranducer\/} is an eight-tuple $(Q,\Sigma,\Gamma,
 \Delta,\delta,q_0,Z_0,F)$, where all symbols have the same meaning as for a
 pushdown automaton except that $\Delta$ is a finite {\em output alphabet\/}
 and $\delta$ is now a mapping $\delta: Q \times (\Sigma \cup \{\varepsilon\})
 \times \Gamma \mapsto Q \times \Gamma^* \times \Delta^*$.

 For a context-free grammar $G = (V,\Sigma,P,S)$, an integer $k$, and $\alpha 
 \in V^*$ the set $FIRST_k(\alpha)$ contains all terminal strings of length
 $\leq k$ and all prefixes of length $k$ of terminal strings which can be 
 derived from $\alpha$ in $G$. More formally,
 $$FIRST_k(\alpha) := \{x \in \Sigma^* \mid  \alpha \stackrel{*}{\Rightarrow} xy, \; y \in 
 \Sigma^+ \mbox{ and } |x| =  k  \mbox{ or } y = \varepsilon \mbox{ and } |x| \leq k\}.
$$
 We will use efficient data structures for the representation of $FIRST_k$-sets.
 A usual way to represent a finite set of strings is the use of a trie. Let
 $\Sigma$ be a finite alphabet of size $l$. A {\em trie\/} with 
 respect to $\Sigma$ is a directed tree $T = (V,E)$ where each node $v \in V$ 
 has outdegree $\leq l$.
 The outgoing edges of a node $v$ are marked by pairwise distinct elements of
 the alphabet $\Sigma$. The node $v$ represents the string $s(v)$ which is
 obtained by the concatenation of the edge markings on the unique path from the 
 root $r$ of $T$ to $v$. 
 An efficient algorithm without the use of fixed-point iteration for the 
 computation of all $FIRST_k$-sets can be found in \cite{Bl1}.

 Let $G =(V,\Sigma,P,S)$ be a reduced, context-free grammar and $k \geq 0$ be an 
 integer. We say that $G$ is $LR(k)$ if
\vspace{-0.15cm}
 \begin{enumerate}
\item
$S \stackrel{*}{\Rightarrow} \alpha Aw \Rightarrow \alpha\beta w$, 
\item
$S \stackrel{*}{\Rightarrow} \gamma Bx \Rightarrow \alpha\beta y$, and
\item
$FIRST_k(w) = FIRST_k(y)$
 \end{enumerate}
\vspace{-0.15cm}
 imply $\alpha = \gamma$, $A = B$, and $x = y$.

\smallskip
In the next three sections, the necessary background is given. Section 3 describes the
canonical $LR(k)$-parser. Section 4 presents the pushdown automaton $M_G$ designed for
an arbitrary context-free grammar $G$. Its efficient simulation is described in Section 5. 
Section 6 combines the canonical $LR(k)$-parser and the efficient simulation of $M_G$ for an 
arbitrary $LR(k)$-grammar $G$ obtaining the extended $LR(k)$-parser for $G$. In Section 7, 
an efficient implementation of the $LR(k)$-parser is described. Section 8 presents some 
experimental results. 

\section{The Canonical $LR(k)$-Parser}

 For the construction of a parser for an $LR(k)$-grammar $G$ the following notations
 are useful:
 Let $S \stackrel{*}{\Rightarrow} \alpha Aw \Rightarrow \alpha\beta w$ be a rightmost 
 derivation in $G$. A prefix $\gamma$ of $\alpha\beta$ is called {\em viable prefix\/}
 of $G$.
 A production in $P$ with a dot on its right side is an {\em item}. More
 exactly, let $p = X \rightarrow X_1X_2 \ldots X_{n_p} \in P$. Then $[p,i]$,
 $0 \leq i \leq n_p$ is an {\em item} which is represented by $[X \rightarrow 
 X_1X_2 \ldots X_i \cdot X_{i+1} \ldots X_{n_p}]$. If $p = X \rightarrow \varepsilon$ then we
simply write $[X \rightarrow \cdot]$. If we add to an item a terminal
 string of length $\leq k$ then we obtain an $LR(k)$-item. More formally, 
 $[A \rightarrow \beta_1 \cdot \beta_2,u]$ where $A \rightarrow \beta_1\beta_2 \in P$ and 
$u \in \Sigma^{\leq k}$ is an {\em $LR(k)$-item\/}. An $LR(k)$-item
 $[A \rightarrow \beta_1 \cdot \beta_2,u]$ is {\em valid\/} for $\alpha\beta_1 \in
 V^*$ if there is a derivation $S \stackrel{*}{\Rightarrow} \alpha Aw \Rightarrow
 \alpha\beta_1\beta_2w$ with $u \in FIRST_k(\beta_2w)$. 
 Note that by definition, an $LR(k)$-item can only  be valid for a viable prefix of
 $G$.

 \smallskip
 The canonical $LR$-parser is a {\em shift-reduce parser\/}. A shift-reduce parser is
 a pushdown automaton which constructs a rightmost derivation backwards. We will give 
 an informal description of such a pushdown automaton. Let
 $S \Rightarrow \alpha_0 \Rightarrow \alpha_1 \Rightarrow \ldots \Rightarrow \alpha_{m-1} 
 \Rightarrow \alpha_m = x$
 be a rightmost derivation of $x$ from $S$. The shift-reduce parser starts with the
 right sentential form $\alpha_m := x$ as input and constructs successively the right
 sentential forms $\alpha_{m-1}, \alpha_{m-2}, \ldots , \alpha_1, \alpha_0, S$. The
 current right sentential form will always be the concatenation of the content of the
 pushdown store from the bottom to the top and the unread suffix of the input.
 At the beginning, the pushdown store is empty. Let $y$ be the unexpended input and
 $\alpha_i = \gamma y$ be the current right sentential form. Then $\gamma$ is the 
 current content of the pushdown store where the last symbol of $\gamma$ is the
 uppermost symbol of the pushdown store. Our goal is to construct the right sentential form 
$\alpha_{i-1}$ from $\alpha_i$.

 If $\alpha_i = \gamma_1\gamma_2 y$ and $\alpha_{i-1} = \gamma_1 Ay$ then the 
 alternative $\gamma_2$ of the variable $A$ expanded in the current step is on the
 top of the stack. If $\alpha_i = \gamma_1\gamma_2 y_1y_2$ and $\alpha_{i-1} = 
 \gamma_1 Ay_2$ then a portion of the alternative of $A$ is prefix of the unexpended 
 input $y$. The goal of the shift-reduce parser is to take care that the alternative 
 of the variable $A$ expanded in 
 $\alpha_{i-1}$ is on the top of the stack. If the alternative of $A$ is on the top of
 the stack then the shift-reduce parser replaces this alternative by $A$. For doing
 this, the shift-reduce parser uses the following operations:
 \begin{enumerate}
 \item
 The next input symbol is read and shifted on the top of the pushdown store.
 \item
 The shift-reduce parser identifies that the alternative of $A$ is on the top of the 
 stack and replaces this alternative by $A$. Therefore, a reduction is performed.
 \end{enumerate}
 In each step, the shift-reduce parser can perform any of the two operations. In 
 general, the shift-reduce parser is nondeterministic. $LR(k)$-grammars allow
 to make the shift-reduce parser deterministically. Moreover, the set of the 
 $LR(k)$-items valid for the current content of the stack contains always the 
 information which is
 sufficient to decide uniquely the next step of the shift-reduce parser. For the
 proof of this central theorem we need the following lemma which gives a more specific 
 characterization of context-free grammers which are not $LR(k)$. 
 \begin{lem}
 Let $k \geq 0$ be an integer and $G = (V,\Sigma,P,S)$ be a reduced cfg which is not 
 $LR(k)$. Then there exists derivations
\vspace{-0.2cm}
 \begin{enumerate}
 \item
 $S \stackrel{*}{\Rightarrow} \alpha Aw \Rightarrow \alpha\beta w$ \hspace{0.5cm} and
 \item
 $S \stackrel{*}{\Rightarrow} \gamma Bx \Rightarrow \gamma\delta x = \alpha\beta y$
 \end{enumerate}
\vspace{-0.2cm}
 where $FIRST_k(w) = FIRST_k(y)$ and $|\gamma\delta| \geq |\alpha\beta|$ but 
 $\alpha Ay \not= \gamma Bx$.
 \end{lem}
 The proof can be found in \cite{AhUl} (proof of Lemma 5.2 at page 382).  

 \smallskip
 Let $\gamma y$ be the current right sentential form; i.e., $\gamma$ is the current 
 content of the stack and $y$ is the unread suffix of the input. Let 
 $u := FIRST_k(y)$ be the current {\em lookahead\/}. Let $[A \rightarrow \beta_1 \cdot 
 \beta_2,v]$ be an $LR(k)$-item which is valid for $\gamma$. Then $\beta_1$ is a suffix
 of $\gamma$. If $\beta_2 = \varepsilon$ then we have $v = u$ and the $LR(k)$-item 
 $[A \rightarrow \beta_1\cdot,u]$ corresponds to a reduction which can be performed by
 the shift-reduce parser.
 If $\beta_2 \in \Sigma V^*$ and $u \in FIRST_k(\beta_2v)$ then the $LR(k)$-item
 $[A \rightarrow \beta_1\cdot\beta_2,v]$ corresponds to a reading which can be
 performed by the shift-reduce parser. The following theorem tells us that the set of
 all $LR(k)$-items valid for $\gamma$ corresponds to at most one step which can be 
 performed by the shift-reduce parser.
Note that Theorem 1 is a weaker version of Theorem 5.9 in \cite{AhUl} which uses the so-called
$\varepsilon$-free first function $EFF_k(\alpha)$. Since the weaker version suffices such that
the $\varepsilon$-free first function is not needed, we present the theorem and the proof here.
 \begin{theo}
 Let $k \geq 0$ be an integer and $G = (V,\Sigma,P,S)$ be a context-free grammar. $G$
 is $LR(k)$ if and only if for all $u \in \Sigma^{\leq k}$ and all $\alpha\beta \in
 V^*$ the following property is fulfilled: 
 If an $LR(k)$-item $[A \rightarrow \beta\cdot,u]$ is valid for $\alpha\beta$ then 
 there exists no other $LR(k)$-item $[C \rightarrow \beta_1\cdot\beta_2,v]$ valid for 
 $\alpha\beta$ with $\beta_2 \in \Sigma V^* \cup \{\varepsilon\}$ and $u \in 
 FIRST_k(\beta_2v)$. 
 \end{theo}
 {\bf Proof:}
    Assume that $G$ is an $LR(k)$-grammar.
 Let $[A \rightarrow \beta\cdot,u]$ and $[C \rightarrow \beta_1\cdot\beta_2,v]$ with 
 $\beta_2 \in \Sigma V^* \cup \{\varepsilon\}$ and $u \in FIRST_k(\beta_2v)$ be two 
 distinct $LR(k)$-items valid for $\alpha\beta$. Then there are derivations 
 \begin{quote}

 $S \stackrel{*}{\Rightarrow} \alpha A w \Rightarrow \alpha\beta w$ \hspace{0.5cm} and \hspace{0.5cm}
 $S \stackrel{*}{\Rightarrow} \alpha_1 C x \Rightarrow \alpha_1\beta_1\beta_2 x$

 \end{quote}
 where $u = FIRST_k(w)$ and $v = FIRST_k(x)$. 
 We have to prove that $G$ cannot be $LR(k)$. With respect to $\beta_2$
 three cases are possible:

 \smallskip
 \noindent
 {\em Case 1:\/} $\beta_2 = \varepsilon$.

 \smallskip
 Then we have $u = v$. Hence, both derivations look as follows:
 \begin{quote}

 $S \stackrel{*}{\Rightarrow} \alpha A w \Rightarrow \alpha\beta w$ \hspace{0.5cm} and \hspace{0.5cm}
 $S \stackrel{*}{\Rightarrow} \alpha_1 C x \Rightarrow \alpha_1\beta_1 x = \alpha\beta x$,

 \end{quote}
 where $FIRST_k(w) = FIRST_k(x) = u$.
 Since both $LR(k)$-items are distinct, we obtain $A \not= C$ or $\beta \not= \beta_1$.
 Note that $\beta \not= \beta_1$ implies $\alpha \not= \alpha_1$. In both cases, we 
 obtain a contradiction to the definition of $LR(k)$-grammers.

 \smallskip
 \noindent
 {\em Case 2:\/} $\beta_2 = z$ where $z \in \Sigma^+$.

 \smallskip
 Then both derivations look as follows:
 \begin{quote}

 $S \stackrel{*}{\Rightarrow} \alpha A w \Rightarrow \alpha\beta w$ \hspace{0.5cm} and \hspace{0.5cm}
 $S \stackrel{*}{\Rightarrow} \alpha_1 C x \Rightarrow \alpha_1\beta_1 zx$,

 \end{quote}
 where $\alpha\beta = \alpha_1\beta_1$ and $FIRST_k(w) = FIRST_k(zx) = u$.
 $z \in \Sigma^+$ implies that $x \not= zx$. Hence, by the definition of 
 $LR(k)$-grammers, $G$ cannot be $LR(k)$.

 \smallskip
 \noindent
 {\em Case 3:\/} $\beta_2 = a\beta'_2$ where $a \in \Sigma$ and $\beta'_2 \in 
 V^*NV^*$.

 \smallskip
 Then there are $u_1,u_2,u_3$ $\in \Sigma^*$ such that
 $\beta_2 = a\beta'_2 \stackrel{*}{\Rightarrow} au_1Bu_3 \Rightarrow au_1u_2u_3$ with
 $FIRST_k(au_1u_2u_3x) = u$. Therefore, both derivations look as follows:
 \begin{quote}

 $S \stackrel{*}{\Rightarrow} \alpha A w \Rightarrow \alpha\beta w$ \hspace{0.5cm} and \\
 $S \stackrel{*}{\Rightarrow} \alpha_1 C x \Rightarrow \alpha_1\beta_1a\beta'_2x
 \stackrel{*}{\Rightarrow} \alpha_1\beta_1au_1Bu_3x \Rightarrow 
 \alpha_1\beta_1au_1u_2u_3x$

 \end{quote}
 where $\alpha\beta = \alpha_1\beta_1$. Applying the definition of $LR(k)$-grammars to
 the derivations
 \begin{quote}

 $S \stackrel{*}{\Rightarrow} \alpha Aw \Rightarrow \alpha\beta w$ \hspace{0.5cm} and \\
 $S \stackrel{*}{\Rightarrow} \alpha_1\beta_1au_1Bu_3x \Rightarrow \alpha_1\beta_1au_1u_2
 u_3x$,

 \end{quote}
 we obtain $\alpha = \alpha_1\beta_1au_1$. Since $\alpha\beta = \alpha_1\beta_1$ and
 $a \in \Sigma$ this is impossible. Hence, $G$ cannot be $LR(k)$.

 \smallskip
 For the proof of the other direction assume that $G$ is not $LR(k)$. Then Lemma 1
 implies that there are two derivations
 \begin{quote}

 $S \stackrel{*}{\Rightarrow} \alpha A w \Rightarrow \alpha\beta w$ \hspace{0.5cm} and \hspace{0.5cm}
 $S \stackrel{*}{\Rightarrow} \gamma C x \Rightarrow \gamma\delta x = \alpha\beta y$

 \end{quote}
 with $u := FIRST_k(w) = FIRST_k(y), |\gamma\delta| \geq |\alpha\beta|$ and $\alpha Ay
 \not= \gamma Cx$. 
 This implies that the $LR(k)$-item $[A \rightarrow \beta\cdot,u]$ is valid for 
 $\alpha\beta$. Hence, it remains to construct a further $LR(k)$-item $[C \rightarrow
 \beta_1\cdot\beta_2,v]$ with $\beta_2 \in \Sigma V^* \cup \{\varepsilon\}$ and
 $u \in FIRST_k(\beta_2v)$ which is valid for $\alpha\beta$.

 Since $|\gamma\delta| \geq |\alpha\beta|$ and $\gamma\delta x = \alpha\beta y$ there
 holds $\gamma\delta = \alpha\beta z$ for a $z \in \Sigma^*$. Two cases are possible: 
 $z$ is a suffix of $\delta$ or $\delta$ is a suffix of $z$. 

 \smallskip
 If $\delta = \delta' z$ for a $\delta' \in V^*$ then $y = zx$ implies that the
 $LR(k)$-item $[C \rightarrow \delta'\cdot z,v]$ with $v = FIRST_k(x)$ is valid for 
 $\alpha\beta$.
 Assume that $[C \rightarrow \delta'\cdot z,v] = [A \rightarrow \beta\cdot,u]$.
 Then $A = C$, $\beta = \delta'$, $z = \varepsilon$ and $u = v$. This implies also
 $\beta = \delta$ and therefore $\alpha = \gamma$ and $x = y$. But this is a 
 contradiction to $\alpha Ay \not= \gamma Cx$. Hence, 
 $[C \rightarrow \delta'\cdot z,v] \not= [A \rightarrow \beta\cdot,u]$.

 \smallskip
 If $z = z'\delta$ for a $z' \in \Sigma^+$ then we consider the rightmost derivation
 $S \stackrel{*}{\Rightarrow} \alpha\beta z'Cx$. Let $\alpha_1By_1$ be the last right 
 sentential form of this derivation with $y_1 \in \Sigma^*$, $B \in N$ and
 $|\alpha_1B| \leq |\alpha\beta|+1$. Note that $|S| = 1$ implies that this right 
 sentential form exists. Hence, we can write the derivation of $\alpha\beta y$ from 
 $S$ in the following form:
 \begin{quote}

 $S \stackrel{*}{\Rightarrow} \alpha_1By_1 \Rightarrow \alpha_1\beta_1\beta_2y_1
 \stackrel{*}{\Rightarrow} \alpha_1\beta_1y$

 \end{quote}
 where $\alpha_1\beta_1 = \alpha\beta$.  By the choice of the right sentential form
 $\alpha_1By_1$ there holds $|\alpha_1| \leq |\alpha\beta|$ and $\beta_2 \in \Sigma 
 V^*$. Hence, the $LR(k)$-item $[B \rightarrow \beta_1\cdot\beta_2,v]$ where
 $v = FIRST_k(y_1)$ is valid for
 $\alpha\beta$. $FIRST_k(y) =u$ implies $u \in FIRST_k(\beta_2v)$. Since $\beta_2 \in
 \Sigma V^*$ the items $[A \rightarrow \beta\cdot,u]$ and $[B \rightarrow 
 \beta_1\cdot\beta_2,v]$ have to be distinct.
 \QED

 \medskip
 With respect to the shift-reduce parser, the theorem has the following implication:
 If during the construction of the rightmost derivation an $LR(k)$-item 
 $[A \rightarrow \beta\cdot,u]$ is valid for the current content $\gamma$ of the 
 stack and $u$ is the current lookahead then $\beta$ is on the top of the pushdown 
 store and the reduction corresponding to the production $A \rightarrow \beta$ is the 
 only applicable step of the shift-reduce parser. If an $LR(k)$-item $[C \rightarrow 
 \beta_1\cdot\beta_2,v]$ with $\beta_2 \in \Sigma V^*$ is valid for the current 
 content of the stack and the current lookahead $u$ is in $FIRST_k(\beta_2v)$ then the
 reading which corresponds to the first symbol of $u$ is the only applicable step
 of the shift-reduce parser. 

\section{The Pushdown Automaton $M_G$}

 For the parallel simulation of all potential leftmost derivations we need the
 following pushdown automaton:
 Given any context-free grammar $G = (V,\Sigma,P,S)$, we will construct a
 pushdown automaton $M_G$ with $L(M_G) = L(G)$ which produces a leftmost derivation. 
 For a production $p \in P$, $n_p$ denotes the length of the right side of $p$.
 Let $H_G = \{[p,i] \mid p \in P, 0 \leq i \leq n_p\}$ be the set of all items of $G$.
 Then $M_G = (Q,\Sigma,\Gamma,\delta,q_0,Z_0,F)$ is defined by
 \begin{quote}
 $Q = H_G \cup \{[S' \rightarrow \cdot S], [S' \rightarrow S\cdot]\}$, \\
 $q_0 = [S' \rightarrow \cdot S]$, $F = \{[S' \rightarrow S\cdot]\}$,\\
 $\Gamma = Q \cup \{\perp\}$, $Z_0 = \perp$, and \\
 $\delta : Q \times (\Sigma \cup \{\varepsilon\}) \times \Gamma \mapsto 
 2^{Q \times \Gamma^*}$.
 \end{quote}
 $\delta$ will be defined such that $M_G$ simulates a leftmost derivation. With 
 respect to $\delta$, we distinguish three types of steps.
 \begin{itemize}
 \item[(E)]
 {\em expansion\/} 

\smallskip
$\delta([X \rightarrow \beta \cdot A\gamma],\varepsilon,Z) =
  \{([A \rightarrow \cdot\alpha], [X \rightarrow \beta \cdot A\gamma]Z) \mid A 
 \rightarrow \alpha \in P\}.$

\smallskip
\noindent
 The leftmost variable in the left sentential form is replaced by one of its
 alternatives. The pushdown store is expanded.
 \item[(C)]
 {\em reading\/}

\smallskip 
$\delta([X \rightarrow \varphi \cdot a\psi],a,Z) = \{([X \rightarrow \varphi 
 a \cdot \psi],Z)\}$.

\smallskip
\noindent
 The next input symbol is read.
 \item[(R)]
 {\em reduction\/}
 
\smallskip
$\delta([X \rightarrow \alpha\cdot],\varepsilon,
 [W \rightarrow \mu \cdot X\nu])
 = \{([W \rightarrow \mu X \cdot\nu],\varepsilon)\}$.

\smallskip
\noindent
 The whole alternative $\alpha$ is derived from $X$. Hence, the dot can be moved
 beyond $X$ and the corresponding item can be removed from the pushdown store
 getting the new state. Therefore, the pushdown store is reduced.
 \end{itemize}
 The basis for the construction of a polynomial size extended $LR(k)$-parser is an
 efficient deterministic simulation of $M_G$. This simulation will be described in
 the next section.

\section{The Deterministic Simulation of $M_G$} 

Let $G = (V,\Sigma,P,S)$ be a reduced context-free grammar.
Our goal is to develop a deterministic simulation of the pushdown automaton $M_G$ 
defined in the previous section. The algorithm which we will develop looks much like Earley's 
algorithm \cite{Ea}. But in contrast to Earley's algorithm, the algorithm 
maintains the structure of the computation of the underlying pushdown automaton $M_G$.
For the construction of the extended $LR(k)$-parser, this structure of the computation of $M_G$
is needed. 
Tomita \cite{To} has develloped a similiar approach the ``graph-structured stack'' which is 
restricted to non-cyclic grammars such that the graphs remain to be acyclic. Next we will describe 
the simulation of $M_G$.

\smallskip
If we write the current state of $M_G$ always on the
top of the stack then we have only to solve the problem of the deterministic
simulation of the stack. The idea is to simulate all possible contents of the stack 
in parallel. Since an exponential number of different stacks are possible at the 
same time, the direct simulation of all stacks in parallel cannot be
efficient. Observe that the grammar $G$ and therefore the pushdown automaton $M_G$
have a fixed size. Hence, at any time, at most a constant number of distinct items
can be on the top of all stacks. Hence, there are only a constant number of 
possibilities to modify eventually an exponential number of different stacks. This
observation suggests the following method:

\smallskip
We realize all stacks simultaneously by a directed graph ${\cal G} = ({\cal V},
{\cal E})$. Each node of the graph is marked by an item. We identify each node with 
 its item. The graph contains exactly one node with indegree zero. This node is
 marked by the item $[S' \rightarrow \cdot S]$. We call this node the {\em start node\/} 
 and nodes with outdegree zero {\em end nodes\/}. Everytime, we have a bijection of 
 the paths from the start node to an end node and the possible contents of the 
 stack.
 The algorithm separates into {\em phases\/}. During each phase, we treat all end nodes
 simultaneously. For doing this, we have the difficulty that with respect to different
 end nodes the kind of steps which have to be performed might be different; i.e., some
 end nodes have to be expanded, other end nodes have to be reduced, and some end nodes
 need a reading. Hence, it can be the case that with respect to different end 
 nodes the unexpended input might be different. For the solution of this difficulty,
 we synchronize the computation using the following rules:
 \begin{enumerate}
\item
 As long as there is an end node of the form $[A \rightarrow \alpha_1\cdot B\alpha_2]$,
 $B \in N$ perform an expansion with respect to this end node.
\item
 If all end nodes are of the form $[A \rightarrow \alpha_1\cdot\alpha_2]$, $\alpha_2 \in
 \Sigma V^* \cup \{\varepsilon\}$ then perform a reduction with respect to all end nodes
 with $\alpha_2 = \varepsilon$.
\item
 If all end nodes are of the form $[A \rightarrow \alpha_1\cdot a\alpha_2]$, $a \in \Sigma$ 
 then perform a reading with respect to all end nodes.
 \end{enumerate}
 At the end of each phase exactly one input symbol has been read. Hence, we have $n$ 
 phases where $n$ is the length of the input. We number these phases from 1 to $n$.
 Each phase separates into two subphases. During the first subphase, we perform all 
 possible expansions and reductions. An end node of the form 
$[A \rightarrow \alpha_1\cdot\alpha_2]$ with $\alpha_2 \in NV^*$ is called {\em expansible\/}, 
with $\alpha_2 \in \Sigma V^*$ is called {\em readable\/}, and with $\alpha_2 = \varepsilon$
 is called {\em reducible\/}.

 The first subphase is separated into rounds. 
 In the first round, we perform as long as possible expansions. We call such a round
 {\em expansion step\/}. The same node is inserted only once. Instead of inserting the same 
 node again, an edge pointing to the node inserted before is created. Since the alternative of 
 an expanded nonterminal can be in $NV^*$, possibly we have to expand the new node again.
 Maybe, some 
 cycles are constructed; e.g., the following chain of expansions would produce a cycle:
 $$[A \rightarrow \alpha_1\cdot B\alpha_2],[B \rightarrow \cdot C\beta_1], [C \rightarrow \cdot
    B \beta_2], [B \rightarrow \cdot C\beta_1].
 $$ 
 In the second round, we perform all possible reductions. Such a round is
 called {\em reduction step\/}. According to the reductions, maybe some further 
 expansions are possible. These are performed during a third round. If the alternative
 of the expanded variable is $\varepsilon$ then this new end node is reducible and 
 causes a reduction. All these reductions are performed in the next round a.s.o.
 New nodes are indexed by the number of the current phase.  A reduction step is performed as 
 follows: 
 We remove all reducible nodes from the graph. Two cases with respect to a direct predecessor
 $u$ of a removed node can arise:
 \begin{enumerate}
 \item
 All its successors are reducible and will be removed. Then $u$ is of Type 1.
 \item
 $u$ has successors which will be not removed. Then $u$ is of Type 2.
 \end{enumerate}
 If $u$ is of Type 1 then the dot of the item $u$ will be moved by one position to 
 the right. The index of $u$ is changed to the number of the current phase. If $u$ is of 
 Type 2 then we copy the node $u$ and all ingoing edges of $u$ and move the dot of the 
 copy $u'$ of $u$ by one position to the right. We index $u'$ by the number of the 
 current phase. Possibly, after moving the dot in $u$ or in $u'$, the node $u$ or $u'$ 
 becomes reducible, expansible, or readable. 

 After the first subphase, all end nodes have a terminal symbol behind its dot. 
 During the second subphase, we perform the reading step. Assume that the  $(i+1)$st 
 input symbol $a_{i+1}$ is the first unread input symbol. End nodes where the terminal 
 symbol behind the dot is unequal $a_{i+1}$ cannot lead to an accepting computation of 
 the pushdown automaton $M_G$. Hence, they can be removed from the graph. Nodes where 
 all successors are removed can also be removed. 
 In end nodes with the first symbol behind the dot is $a_{i+1}$, we move the dot one 
 position to the right and change the index of the current item to $i+1$. 

The algorithm maintains the following sets: $H$ contains the end nodes of ${\cal G}$
which have to be expanded during the current phase. $K$ contains exactly those nodes
which had been already expanded during the current phase. $R$ contains the reducible
end nodes of ${\cal G}$. $Pr$ contains those nodes which we have to consider since
some direct predecessors have been reduced.
Altogether, we obtain the following algorithm.

 \bigskip
 \noindent
 {\bf Algorithm} {\sc Simulation$(M_G)$} \\
 \noindent {\bf Input:} A reduced cfg $G = (V,\Sigma,P,S)$ and $w = a_1a_2 \ldots a_n \in
 \Sigma^+$. \\
 \noindent {\bf Output:} ``accept'' if $w \in L(G)$ and ``error'' otherwise. \\
\newpage
 \noindent {\bf Method:}
 \begin{tabbing}
 AA \= AA \= AAA \= AA \= AA \= AA \= AA \= AA \= AA \= \kill
 \> (1) \> Initialize ${\cal G} = ({\cal V},{\cal E})$ by 
 ${\cal V} := \{[S' \rightarrow \cdot S]_0\}$ and 
 ${\cal E} := \emptyset$; \\
 \> \> $H := \{[S' \rightarrow \cdot S]_0\}$; \\
 \> \> $K := \emptyset$; $R := \emptyset$; $Pr := \emptyset$; \\ 
 \> \> $exp := 1$; $red := 0$; \\ 
 \> \> $i := 0$; \\ 
\\
 \> (2) \> {\bf while} $i \leq n$ \\
 \> \> {\bf do} \\
 \> \> (ER) \> {\bf while} $exp = 1$ \\
 \> \> \> {\bf do} \\ 
 \> \> \> \> {\bf while} $H \not= \emptyset$ \\
 \> \> \> \> {\bf do} \\
 \> \> \> \> \> Choose any $[A \rightarrow \alpha_1\cdot B\alpha_2]_i \in H$; \\ 
 \> \> \> \> \> $H := H \setminus \{[A \rightarrow \alpha_1\cdot B\alpha_2]_i\}$; \\ 
 \> \> \> \> \> $K := K \cup \{[A \rightarrow \alpha_1 \cdot B\alpha_2]_i\}$; \\
 \> \> \> \> \> {\bf for} each alternative $\beta$ of $B$ \\
 \> \> \> \> \> {\bf do} \\  
 \> \> \> \> \> \> ${\cal V} := {\cal V} \cup \{[B \rightarrow \cdot\beta]_i\}$ \\
 \> \> \> \> \> \> ${\cal E} := {\cal E} \cup \{([A \rightarrow \alpha_1\cdot 
                   B\alpha_2]_i,[B \rightarrow \cdot\beta]_i)\}$; \\
 \> \> \> \> \> \> {\bf if} $\beta = \varepsilon$ \\
 \> \> \> \> \> \> {\bf then} \\
 \> \> \> \> \> \> \> $R := R \cup \{[B \rightarrow \varepsilon\cdot]_i\}$; $red := 1$ \\
 \> \> \> \> \> \> {\bf fi}; \\
 \> \> \> \> \> \> {\bf if} $\beta \in NV^*$ and $[B \rightarrow \cdot\beta]_i 
                \not\in K$ \\
 \> \> \> \> \> \> {\bf then} \\
 \> \> \> \> \> \> \> $H := H \cup \{[B \rightarrow \cdot\beta]_i\}$ \\
 \> \> \> \> \> \> {\bf fi} \\
 \> \> \> \> \> {\bf od}; \\
 \> \> \> \> {\bf od}; \\
 \> \> \> \> $exp := 0$; \\
 \> \> \> \> {\bf while} $red = 1$ \\
 \> \> \> \> {\bf do} \\
 \> \> \> \> \> {\bf while} $R \not= \emptyset$ \\
 \> \> \> \> \> {\bf do} \\
 \> \> \> \> \> \> Choose any $[A \rightarrow \alpha\cdot]_i \in R$; \\
 \> \> \> \> \> \> $R := R \setminus \{[A \rightarrow \alpha\cdot]_i\}$; \\
 \> \> \> \> \> \> $Pr := Pr \cup \{u \in {\cal V} \mid \mbox{ $u$ is a direct }$\\
 \> \> \> \> \> \> \> \> \> $\mbox{predecessor of $[A \rightarrow \alpha\cdot]_i$}\}$; \\
 \> \> \> \> \> {\bf od}; \\
 \> \> \> \> \> $red := 0$; \\
 \> \> \> \> \> {\bf while} $Pr \not= \emptyset$ \\
 \> \> \> \> \> {\bf do} \\
 \> \> \> \> \> \> Choose any $u \in Pr$; \\
 \> \> \> \> \> \> $Pr := Pr \setminus \{u\}$; \\
 \> \> \> \> \> \> {\bf if} $u$ is of Type 1 \\
 \> \> \> \> \> \> {\bf then} \\
 \> \> \> \> \> \> \> move the dot in $u$ one position \\
 \> \> \> \> \> \> \> to the right; \\
 \> \> \> \> \> \> \> change the index of $u$ to $i$; \\
 \> \> \> \> \> \> \> {\bf if} $u$ is expansible \\
 \> \> \> \> \> \> \> {\bf then} \\
 \> \> \> \> \> \> \> \> $H := H \cup \{u\}$; $exp := 1$ \\
 \> \> \> \> \> \> \> {\bf else} \\
 \> \> \> \> \> \> \> \> {\bf if} $u$ is reducible \\
 \> \> \> \> \> \> \> \> {\bf then} \\
 \> \> \> \> \> \> \> \> \> $R := R \cup \{u\}$; $red := 1$  \\
 \> \> \> \> \> \> \> \> {\bf fi} \\
 \> \> \> \> \> \> \> {\bf fi} \\
 \> \> \> \> \> \> {\bf else} \\
 \> \> \> \> \> \> \> copy $u$ with index $i$ and all ingoing \\
 \> \> \> \> \> \> \> edges of $u$; \\
 \> \> \> \> \> \> \> move the dot in the copy $u'$ one \\ 
 \> \> \> \> \> \> \> position to the right; \\
 \> \> \> \> \> \> \> {\bf if} $u'$ is expansible \\
 \> \> \> \> \> \> \> {\bf then} \\
 \> \> \> \> \> \> \> \> $H := H \cup \{u'\}$; $exp := 1$ \\
 \> \> \> \> \> \> \> {\bf else} \\
 \> \> \> \> \> \> \> \> {\bf if} $u'$ is reducible \\
 \> \> \> \> \> \> \> \> {\bf then} \\
 \> \> \> \> \> \> \> \> \> $R := R \cup \{u'\}$; $red := 1$ \\
 \> \> \> \> \> \> \> \> {\bf fi} \\
 \> \> \> \> \> \> \> {\bf fi} \\
 \> \> \> \> \> \> {\bf fi}; \\
 \> \> \> \> \> {\bf od} \\
 \> \> \> \> {\bf od} \\
 \> \> \> {\bf od}; \\
 \> \> (R) \> Delete all end nodes $[A \rightarrow \alpha\cdot a\beta]_i$ with 
                $a \not= a_{i+1}$; \\
 \> \> \> As long as such nodes exists, delete nodes with the \\ 
 \> \> \> property that all successors are removed; \\   
 \> \> \> Replace each end node $[A \rightarrow \alpha\cdot a_{i+1}\beta]_i$ by the \\
 \> \> \> node $[A \rightarrow \alpha a_{i+1}\cdot\beta]_{i+1}$ and modify $H$ and $R$ \\
 \> \> \> as follows; \\
 \> \> \> {\bf if} $\beta \in NV^*$ \\
 \> \> \> {\bf then} \\
 \> \> \> \> $H := H \cup \{[A \rightarrow \alpha a_{i+1}\cdot\beta]_{i+1}\}$; 
             $exp := 1$ \\
 \> \> \> {\bf fi}; \\
 \> \> \> {\bf if} $\beta = \varepsilon$ \\
 \> \> \> {\bf then} \\
 \> \> \> \> $R := R \cup \{[A \rightarrow \alpha a_{i+1}\cdot]_{i+1}\}$; $red := 1$ \\
 \> \> \> {\bf fi}; \\
 \> \> \> $i := i+1$ \\
 \> \> {\bf od}; \\
 \> (3) \> {\bf if} $[S' \rightarrow S\cdot]_n \in V$ \\
 \> \> {\bf then} \\
 \> \> \> output := ``accept'' \\
 \> \> {\bf else} \\
 \> \> \> output := ``error'' \\
 \> \> {\bf fi}. \\
 \end{tabbing}

 Note that the graph ${\cal G} = ({\cal V},{\cal E})$ can contain some cycles.
 The index of an item is equal to the length of the already read input. This index will 
 be helpful for understanding the simulation algorithm and can be omitted in 
 any implementation of the algorithm.
 The correctness of the algorithm follows from the fact that after each performance
 of a phase there is a bijection between the paths from the start node to the end
 nodes and the possible contents of the stack. This can easily be proved by induction. 
 It is also easy to prove that the algorithm {\sc Simulation$(M_G)$} uses $O(n^3)$ 
 time and $O(n^2)$ space where $n$ is the length of the input. If the context-free 
 grammar $G$ is unambiguous, the needed time reduces to $O(n^2)$.
 
\section{The Construction of the Extended $LR(k)$-Parser}

Let $k \geq 0$ be an integer and let $G = (V,\Sigma,P,S)$ be an arbitrary 
$LR(k)$-grammar. The idea is to combine the concept of the shift-reduce parser and the 
deterministic simulation of the pushdown automaton $M_G$. This means that for the 
construction of the extended parser $P_G$ we use $M_G$ under regard of properties of 
$LR(k)$-grammars. Just as for the construction of the canonical $LR(k)$-parser, Theorem 1 is
the key for the construction of the extended $LR(k)$-parser.
Note that Theorem 1 is a statement about valid $LR(k)$-items for a viable prefix
of $G$. Hence, we are interested in all maximal viable prefixes represented by the current 
graph ${\cal G} = ({\cal V},{\cal E})$ of the simulation algorithm of $M_G$. In the 
subsequence, we omit the indices of the items if they are not needed.
Let $[A \rightarrow \alpha_1\cdot\alpha_2]$ be an item. Then we call the portion
$\alpha_1$ left from the dot the {\em left side\/} of the item
$[A \rightarrow \alpha_1\cdot\alpha_2]$.
Let $P$ be any path from the start node to an end node in ${\cal G}$. Then the concatenation
of the left sides of the items from the start node to the end node of $P$ results in the
maximal viable prefix $pref(P)$ with respect to $P$; i.e., if
$$
P = [S' \rightarrow \cdot S],[S \rightarrow \alpha_1\cdot A_2\beta_1],[A_2 
\rightarrow \alpha_2\cdot A_3\beta_2], \ldots,[A_t \rightarrow \alpha_t\cdot\beta_t]
$$
then
$$pref(P) = \alpha_1\alpha_2 \ldots \alpha_t.$$
Next we will characterize valid $LR(k)$-items with respect to such a path $P$ where the end
node of $P$ is reducible or readable; i.e., $\beta_t = \varepsilon$ or $\beta_t = a\beta_t'$
where $a \in \Sigma$ and $\beta_t' \in V^*$. 
Let $[B \rightarrow \alpha\cdot C\beta]$, $C \in N$, $\beta \in V^*$ be an item. Then we call 
$\beta$ the {\em right side\/} of the item $[B \rightarrow \alpha\cdot C\beta]$. 
The {\em right side\/} of an item $[B \rightarrow \alpha\cdot a\beta]$, $a \in 
\Sigma$, $\beta \in V^*$ is $a\beta$. 
We obtain the {\em relevant suffix\/} $suf(P)$ with respect to $P$ by 
concatenating the right sides from the end node to the start node of $P$; i.e., 
$$
suf(P) =  \left\{ \begin{array}{ll}
\beta_{t-1}\beta_{t-2} \ldots \beta_1 & \mbox{if } \beta_t = \varepsilon \\
a\beta_{t}'\beta_{t-1}\beta_{t-2} \ldots \beta_1  & \mbox{if } \beta_t = a\beta_t'.
\end{array}
\right.
$$
Let $u$ be the current lookahead. The $LR(k)$-item $[A_t \rightarrow \alpha_t\cdot\beta_t, u]$
is {\em valid for the path $P$} iff $u \in FIRST_k(suf(P))$. 

\smallskip
For an application of Theorem 1 to $M_G$ it would be useful if all maximal viable prefixes
of a path corresponding to any current stack would be the same. Let us assume for a
moment that this would be the case. Then we can incorporate Theorem 1 into the pushdown 
automaton $M_G$. We call the resulting pushdown automaton $LR(k)$-$M_G$. During the deterministic
simulation of $LR(k)$-$M_G$ the following invariant will be fulfilled:
\begin{enumerate}
\item
Immediately before an expansion step, all end nodes of the graph ${\cal G}$ are of the form 
$[A \rightarrow \alpha\cdot B\beta]$ or $[A \rightarrow \alpha\cdot a\beta]$ where
$\alpha,\beta \in V^*$, $B \in N$ and $a \in \Sigma$ is the next unread symbol of the input.
\item
Immediately before a reduction/reading step, all end nodes of ${\cal G}$ are of the form
$[A \rightarrow \cdot]$ or $[A \rightarrow \alpha\cdot a\beta]$ where
$\alpha,\beta \in V^*$ and $a \in \Sigma$ is the next unread symbol of the input.
\end{enumerate}
Before the first expansion step, the only node of ${\cal G}$ is the start node 
$[S' \rightarrow \cdot S]$. Hence, the invariant is fulfilled before the first expansion step.
Assume that the invariant is fulfilled before the current expansion step. Let
$a$ be the next unread symbol of the input. Since an alternative $\alpha \in 
(\Sigma\setminus\{a\}) V^*$ cannot lead to an accepting computation, all possible expansions are 
performed under the restriction that only alternatives in $NV^* \cup \{a\}V^* \cup 
\{\varepsilon\}$ are used. If a variable $C$ of an end node $[B \rightarrow \alpha\cdot C\beta]$ 
has only alternatives in $(\Sigma\setminus\{a\})V^*$ then this end node cannot lead to an 
accepting computation. Hence, such an end node is removed.
Then, a {\em graph adjustment\/} is performed; i.e., 
as long as there is a node where all its successors are removed from the graph ${\cal G}$
this node is removed, too. Obviously, the invariant is fulfilled after the expansion step and
hence, before the next reduction/reading step.

\smallskip
Assume that the invariant is fulfilled before the current reduction/reading step.
Let $u$ be the current lookahead. Three cases can arise: 

\smallskip
\noindent
{\em Case 1:\/}
There is a path $P$ from the start node to an end node $[A \rightarrow \cdot]$ such 
that the $LR(k)$-item $[A \rightarrow \cdot,u]$ is valid for $P$.

\smallskip
Then according to Theorem 1, the corresponding reduction is the unique step performed by the 
parser. Hence, all other end nodes of the graph ${\cal G}$ are removed. 
Then a graph adjustment and the reduction with respect to the end node $[A \rightarrow \cdot]$ 
are performed. For 
each direct predecessor $u$ of the node $[A \rightarrow \cdot]$ which has Type 1, 
we move the dot of the item $u$ one position to the right. If $u$ is of Type 2 then we copy
$u$ and all ingoing edges and move the dot of the copy $u'$ one position to the right.
The resulting items are the new end nodes of the graph and are of the form 
$[B \rightarrow \alpha A\cdot\beta]$ where $\beta \in V^*$. 
If $\beta \in N V^*$ then the item $[B \rightarrow \alpha A \cdot\beta]$ is expansible.
If $\beta = \varepsilon$ then the resulting item $[B \rightarrow \alpha A\cdot]$ is reducible. 
If the $LR(k)$-item $[B \rightarrow \alpha A\cdot,u]$ is valid then the reduction with 
respect to the end node $[B \rightarrow \alpha A\cdot]$ is performed.
Since after the expansion of any expansible end node each constructed reducible end node would 
be of the form $[C \rightarrow \cdot]$, by Theorem 1, all constructed end nodes cannot
lead to a valid $LR(k)$-item. Hence, we need not perform the expansion of any expansible end
node if the reduction of the end node $[B \rightarrow \alpha A\cdot]$ is performed such that
all other end nodes are removed from the graph. 
If the $LR(k)$-item $[B \rightarrow \alpha 
A\cdot,u]$ is not valid then the end node $[B \rightarrow \alpha A\cdot]$ is removed. 
If $\beta \in (\Sigma\setminus \{a\})V^*$ then 
this end node cannot lead to an accepting computation and can be removed from the graph. 
Then, a graph adjustment is performed. Hence, after the 
reduction step, all end nodes are of the form $[B \rightarrow \alpha A\cdot\beta]$ with 
$\beta \in NV^* \cup \{a\}V^*$. Therefore, the invariant is fulfilled before the next step.

\smallskip
\noindent
{\em Case 2:\/}
There is no such a path $P$ but there is at least one end node with the terminal symbol $a$ 
behind the dot. 

\smallskip
Then, the corresponding reading step is the only possible step performed by the 
parser. All end nodes which do not correspond to this reading step are removed from the graph 
followed by a graph adjustment.
Then we perform the reading step with respect to all remaining end nodes. This means that 
the next input symbol $a$ is read and the dot is moved one position to the right with respect to 
all end nodes. Then the resulting items are of the form $[B \rightarrow \alpha a\cdot\beta]$
where $\beta \in V^*$.
Let $a'$ be the next unread input symbol and $u'$ be the current lookahead.
The same discussion as above shows that after the termination of the current reduction/reading 
step all end nodes are of the form $[B \rightarrow \alpha a\cdot\beta]$ with $\beta \in NV^* 
\cup \{a'\}V^*$. Hence, the invariant is fulfilled before the next step. 

\smallskip
\noindent
{\em Case 3:\/}
None of the two cases described above is fulfilled. 

\smallskip
Then, the $LR(k)$-grammar $G$ does not generate the input.

\smallskip
The following lemma shows that the same maximal viable prefix corresponds to all paths 
from the start node to an end node in ${\cal G}$. Hence, Theorem 1 can be 
applied during each reduction/reading step.
\begin{lem}
Let $G = (V,\Sigma,P,S)$ be an $LR(k)$-grammar and let ${\cal G} = ({\cal V},
{\cal E})$ be the graph constructed by the deterministic simulation of $LR(k)$-$M_G$. 
Then at any time for any two paths $P$ and $Q$ from the start node $[S' \rightarrow \cdot S]$ 
to any end node it holds $pref(P) = pref(Q)$.
\end{lem}
{\bf Proof:}
  We prove the lemma by induction on the number of performed reductions and readings.
At the beginning, the only node of the graph is the start node 
$[S' \rightarrow \cdot S]$. An expansion does not change the maximal viable prefix with 
respect to any path 
since the left side of any corresponding item is $\varepsilon$. Hence, after the first
round of the first subphase, $\varepsilon$ is the unique maximal viable prefix of all 
paths from the start to an end node of ${\cal G}$. This implies that the assertion 
holds before the first reduction or reading.

Assume that the assertion is fulfilled after $l$, $l \geq 0$, 
reductions/readings. The expansions performed between the $l$th and the $(l+1)$st
reduction/reading do not change the maximal viable prefix of any
path from the start to an end node of ${\cal G}$. Hence, the assertion is fulfilled 
immediately before the $(l+1)$st reduction/reading. 
Let $\gamma$ be the unique maximal viable prefix which corresponds to all paths from
the start to an end node of the graph ${\cal G}$. Let $y$ be the unread suffix of the
input. Two cases can arise:

 \smallskip
 \noindent
 {\em Case 1:\/} A reduction is applicable.

 \smallskip
 Then the maximal viable prefix $\gamma$ has the form $\gamma = \alpha\beta$ such that
 there is an $LR(k)$-item $[A \rightarrow \beta\cdot, FIRST_k(y)]$ which is valid with respect
 to a path $P$ from the start node to an end node $[A \rightarrow \beta\cdot]$.
 Theorem 1 implies that no other $LR(k)$-item $[C \rightarrow
 \beta_1\cdot\beta_2,v]$ with $\beta_2 \in \Sigma V^* \cup \{\varepsilon\}$ is valid 
 for $\gamma$. Hence, no other reduction and no reading is applicable.
 All end nodes which are not consistent with the only applicable reduction are removed 
 from the graph. As long as there is a node such that all its successors are removed 
 from the graph this node is removed, too.
 For the remaining end node $[A \rightarrow \beta\cdot]$ the reduction
 is performed. This implies that the maximal viable prefix of all corresponding paths 
 from the start to an end node of ${\cal G}$ is $\alpha A$.
 Altogether, after the $(l+1)$st reduction/reading all paths from the start to an end node 
 have the maximal viable prefix $\alpha A$.

 \smallskip
 \noindent
 {\em Case 2:\/} No reduction is applicable.

 \smallskip
 Let $y = ay'$ where $a \in \Sigma$, $y' \in \Sigma^*$. If ${\cal V} = \emptyset$ then the input 
is not in the language generated by the $LR(k)$-grammar $G$. Otherwise, all end nodes of the 
current graph ${\cal G}$ have the terminal symbol $a$ behind the dot. Moreover, all paths from
 the start node to an end node have the maximal viable prefix $\gamma$.
 We perform with respect to all end nodes of the current graph the reading of the next unread 
input symbol. That is
 the dot is moved one position to the right behind $a$. After doing this, all paths from
 the start node to an end node have the maximal viable prefix $\gamma a$.
 
 \smallskip
 Altogether, after the $(l+1)$st reduction/reading, the assertion is fulfilled.
 \QED

\section{The Implementation of the Simulation of $LR(k)$-$M_G$}

How to realize the implementation of the simulation of $LR(k)$-$M_G$ described above? Mainly,
the following questions arise:
\begin{enumerate}
\item
How to perform the expansions efficiently?
\item
How to perform a reduction/reading step efficiently?
\end{enumerate}
Let $i$ be the index of the current phase and $a$ be the next unread input symbol.
Assume that $\gamma_1, \gamma_2, \ldots,\gamma_q$ are those alternatives of the variable $A$
which are in $NV^* \cup \{a\}V^* \cup \{\varepsilon\}$.
The expansion of an end node $[C \rightarrow \alpha\cdot A\beta]_i$ of the current graph 
${\cal G}$ is performed in the following way:
\begin{itemize}
\item[(1)]
If the variable $A$ is expanded during the current phase for the first time then 
add the nodes $[A \rightarrow \gamma_j]_i$, $1 \leq j \leq q$ to the current graph ${\cal G}$.
\item[(2)]
Add the edges $([C \rightarrow \alpha\cdot A\beta]_i,[A \rightarrow \cdot\gamma_j]_i)$,
$1 \leq j \leq q$ to the current graph ${\cal G}$.
\end{itemize}
If the variable $A$ is expanded for the first time then $q$ nodes and $q$ edges are added to the 
graph. If after this expansion another end node
$[C' \rightarrow \alpha'\cdot A\beta']_i$ has to be expanded we would add the $q$ edges
$([C' \rightarrow \alpha'\cdot A\beta']_i,[A \rightarrow \cdot\gamma_j]_i)$ to the graph.
Therefore, the number of nodes of the graph ${\cal G}$ is bounded by $O(|G|n)$ but the number
of edges can increase to $O(|G|^2n)$. Hence, our goal is to reduce the number of edges in 
${\cal G}$. The idea is to create an additional node $A_i$ and the edges 
$(A_i,[A \rightarrow \cdot\gamma_j]_i)$, $1 \leq j \leq q$. Then, the expansion of an end node 
$[C \rightarrow \alpha\cdot A\beta]_i$ of the current graph ${\cal G}$ can be performed in the 
following way:
\begin{itemize}
\item[(1)]
If the variable $A$ is expanded during the current phase for the first time then add 
the nodes $A_i$ and $[A \rightarrow \gamma_j]_i$, $1 \leq j \leq q$ and the edges 
$(A_i,[A \rightarrow \cdot\gamma_j]_i)$, $1 \leq j \leq q$ to the current graph ${\cal G}$.
\item[(2)]
Add the edge $([C \rightarrow \alpha\cdot A\beta]_i,A_i)$ to the current graph ${\cal G}$.
\end{itemize}
Then $q+1$ edges are inserted for the first expansion of the variable $A$. For each further
expansion of $A$ during the current phase only one edge is inserted. This will lead to 
an $O(|G|n)$ upper bound for the number of edges in ${\cal G}$.
The expansion step transforms the graph ${\cal G}$ to a graph ${\cal G}'$.

\smallskip
After the expansion step, a reduction/reading step has to be performed. Let $u$ be the current
lookahead. First, we check if there is a path $P$ from the start node to an end node 
$[A \rightarrow \cdot]$ or $[A \rightarrow \alpha \cdot a\beta]$  such that 
the $LR(k)$-item $[A \rightarrow \cdot,u]$ and $[A \rightarrow \alpha \cdot a\beta,u]$,
respectively is valid for $P$. We call such a path $P$ {\em suitable} for the end node 
$[A \rightarrow \cdot]$ and $[A \rightarrow \alpha\cdot a\beta]$, respectively.
For doing this, we need an answer to the following question: 
\begin{itemize}
\item[]
Given such an end node $[A \rightarrow \cdot]$ or $[A \rightarrow \alpha\cdot a\beta]$
and a path $P$ from the start node to this end node, how to decide efficiently if 
$u \in FIRST_k(suf(P))$?
\end{itemize}
The complexity of the reduction/reading step mainly depends on the length $k$ of the lookahead $u$.
First, we will describe the implementation of the reduction/reading step for small $k$ and then
for large $k$.

\subsection{Lookaheads of small size}

We will consider the most simple case $k = 1$ first and then larger small lengths. Let
$k = 1$. We distinguish two cases:

\smallskip
\noindent
{\em Case 1:} There is an end node of the form $[A \rightarrow \alpha\cdot a\beta]$ where
              $\alpha,\beta \in V^*$ and $a \in \Sigma$.

\smallskip
According to the invariant which is fulfilled during the simulation of $LR(k)$-$M_G$, the 
terminal symbol $a$ is the next unread symbol of the input.
Obviously, $a \in FIRST_1(suf(P))$ for all paths $P$ from the start node to the end node
$[A \rightarrow \alpha\cdot a\beta]$. Hence, the $LR(k)$-item $[A \rightarrow \alpha\cdot a\beta,a]$
is valid for all such paths. Theorem 1 implies that no $LR(k)$-item which does not correspond 
to reading the next input symbol can be valid for a path from the start node to an end 
node.

\smallskip
\noindent
{\em Case 2:} All end nodes of $\cal{G}'$ are of the form $[A \rightarrow \cdot]$.

\smallskip
Let $P$ be a path from the start node to the end node $[A \rightarrow \cdot]$ and
let $suf(P) = A_1A_2 \ldots A_r$. Then $A_i \in V$, $1 \leq i \leq r$.
The $LR(k)$-item $[A \rightarrow \cdot, a]$ is valid for $P$ iff $a \in 
FIRST_1(A_1A_2 \ldots A_r)$. Note that $a \in FIRST_1(A_1A_2 \ldots A_r)$ iff
$a \in FIRST_1(A_1)$ or $\varepsilon \in FIRST_1(A_1)$ and $a \in FIRST_1(A_2A_3 \ldots A_r)$. 
Hence, $a \in FIRST_1(suf(P))$ iff there is $1 \leq i \leq r$ such that $\varepsilon \in 
FIRST_1(A_1A_2 \ldots A_{i-1})$ and $a \in FIRST_1(A_i)$. 
For the decision if $a \in FIRST_1(suf(P))$, we consider $A_1A_2 \ldots A_r$ from left to right. 
Assume that $A_j$ is the current considered symbol. If $a \in FIRST_1(A_j)$ then 
$a \in FIRST_1(suf(P))$.
Otherwise, if $\varepsilon \not\in FIRST_1(A_j)$ or $j = r$ then $a \not\in FIRST_1(suf(P))$.
If $\varepsilon \in FIRST_1(A_j)$ and $j < r$ then the next symbol $A_{j+1}$ of $suf(P)$ is
considered.

\smallskip
Now we know how to decide if the current lookahead $a$ is contained in $FIRST_1(suf(P))$ for
a given path $P$ from the start node to a readable or reducible end node. But we have to solve 
the following more general problem: 
\begin{itemize}
\item[]
Given an end node $[A \rightarrow \alpha\cdot a\beta]$ or $[A \rightarrow \cdot]$, how 
to decide if there is a path $P$ from the start node to this end node with $a \in FIRST_1(suf(P))$? 
\end{itemize}
The first case is trivial since for all paths $P$ from the start node to the end node 
$[A \rightarrow \alpha\cdot a\beta]$ there holds $a \in FIRST_1(suf(P))$.
In the second case, there can be a large number of paths from the start node to the end node 
$[A \rightarrow \cdot]$ such that we cannot answer this question by checking each 
such a path separately. Hence, we check all such paths simultaneously. The idea is to apply 
an appropriate graph search method to $\cal{G}'$. 

\smallskip
A {\em topological search\/} on a directed graph is a search 
which visits only nodes with the property that all its predecessors are already visited. 
A {\em reversed search\/} on a directed graph is a search on the graph where the edges are 
traversed against their direction.
A {\em reversed topolgical search\/} on a directed graph is a reversed search which visits only 
nodes where all its successors are already visited. Note that topological search and reversed 
topological search can only be applied to acyclic graphs.

\smallskip
It is useful to analyze the structure of the graph ${\cal G}(A)$ which is constructed according 
the expansion of the variable $A$. The graph ${\cal G}(A)$ depends only on the grammar and not on
the current input of the parser. Note that ${\cal G}(A)$ has the unique start node $A$.
The nodes without successors are the end nodes of ${\cal G}(A)$. An expansion step only inserts 
nodes where the left side of the corresponding item is $\varepsilon$. A successor 
$[C \rightarrow \cdot A\beta]$ of the start node $A$ in ${\cal G}(A)$ is called {\em final node\/}
of ${\cal G}(A)$. Observe that $([C \rightarrow \cdot A\beta],A)$ is an edge which closes a cycle 
in ${\cal G}(A)$. We call such an edge {\em closing edge\/}. Such cycles formed by closing edges
are the only cycles in ${\cal G}'$.

\smallskip
The idea is to perform a reversed topological search on ${\cal G}'$ although ${\cal G}'$ is not
acyclic. The following questions have to be answered:
\begin{enumerate}
\item
What is the information which has to be transported through the graph during the search?
\item
How to treat the cycles in ${\cal G}'$ during the reversed topological search?
\end{enumerate}
At the beginning of the reversed topological search, the only nodes where all successors are 
already visited are the end nodes of ${\cal G}'$. Hence, the search starts with an end node.
If the graph ${\cal G}'$ contains a readable end node then the search starts with a readable
end node. Otherwise, the search starts with a reducible end node.
We discuss both cases one after the other.

\smallskip
\noindent
{\em Case 1\/}: There exists a readable end node.

\smallskip
Then $a \in FIRST_1(suf(P))$ for all paths $P$ from the start node to a readable end node. All
reducible end nodes are removed from the graph. As long as there is a node such that all its 
successors are removed from the graph this node is removed, too. In all remaining end nodes
the dot is moved one position to the right. This terminates the current phase.

\smallskip
\noindent
{\em Case 2\/}: There exists no readable end node.

\smallskip
Assume that the search starts with the end node $[A \rightarrow \cdot]$.
Then $a$ has to be derived from right sides of items 
which correspond to predecessors of the node $[A \rightarrow \cdot]$. This information 
associated with the end node $[A \rightarrow \cdot]$ has to be transported to its
direct predecessor $A$.

Nodes which correspond to an item have outdegree zero or one. Only nodes which correspond to
a variable $C$ can have outdegree larger than one. If one visit the node $C$ during the 
backward topological search then we know that there is a path $Q$ with start node $C$ such that
$\varepsilon \in FIRST_1(suf(Q))$. If the node $C$ would be visited over two different outgoing 
edges of $C$ then there would exist at least two different such paths $Q$ and $Q'$.
Since all paths from the start node of $\cal{G}'$ to an end node have the same maximal viable prefix
it follows that $pref(Q) = pref(Q')$. Hence, there would be a word in $L(G)$ having at least two 
different leftmost derivations. Since $LR(k)$-grammers are unambiguous this cannot happen.
Therefore, such a node $C$ can only be visited over one of its outgoing edges during the 
reversed topological search. 

Assume that we enter the node $C$ over an outgoing edge $e$ and there is a closing edge 
$([B \rightarrow \cdot C\gamma],C)$.
Before the continuation of the reversed topological search at the node $C$, the search is 
continued at the node $[B \rightarrow \cdot C\gamma]$. Note that this node is a successor of the 
node $C$ in the graph ${\cal G}'$. But it cannot happen that we visit the node $C$ again since this
would imply that the grammar has to be ambiguous. Hence, either one finds a path $P$ such that
$a \in FIRST_1(suf(P))$ or one detects that no such a path using the cycle exists before reaching
the node $C$ again. This observation implies that, although the graph ${\cal G}'$ may contain some
cycles, for the reversed topological search the graph can be considered as an acyclic graph.
After the treatment of all closing edges with end node $C$, the reversed topological search is 
continued at the node $C$.

\smallskip
Let $P_0$ be a path from the start node to the end node $[A \rightarrow \cdot]$ in
${\cal G}'$ with $a \in FIRST_1(suf(P))$. It is worth to investigate the structure of $P_0$. Let
$$
P_0 = [S' \rightarrow \cdot S], \ldots, [C_t \rightarrow \cdot C_{t-1}\gamma_t], \ldots,
     [C_2 \rightarrow \cdot C_1\gamma_2], [C_1 \rightarrow \cdot A\gamma_1], 
     [A \rightarrow \cdot]
$$
where $\gamma_i \in N^*$, $\varepsilon \in FIRST_1(\gamma_i)$, $a \not\in FIRST_1(\gamma_i)$ for 
$1 \leq i \leq t-1$ and $a \in FIRST_1(\gamma_t)$. After the reduction of the end node
$[A \rightarrow \cdot]$, we obtain from $P_0$ the path
$$
P_1 = [S' \rightarrow \cdot S], \ldots, [C_t \rightarrow \cdot C_{t-1}\gamma_t], \ldots,
     [C_2 \rightarrow \cdot C_1\gamma_2], [C_1 \rightarrow A\cdot\gamma_1]
$$
Then $\varepsilon$ is derived from $\gamma_1$. Since $LR(k)$-grammars are unambiguous, there is
a unique derivation of $\varepsilon$ from $\gamma_1$. After the performance of all corresponding
expansions and reductions, we obtain from $P_1$ the path
$$
P_2 = [S' \rightarrow \cdot S], \ldots, [C_t \rightarrow \cdot C_{t-1}\gamma_t], \ldots,
     [C_2 \rightarrow \cdot C_1\gamma_2], [C_1 \rightarrow A\gamma_1\cdot]. 
$$
Then, the end node $[C_1 \rightarrow A\gamma_1\cdot]$ is reduced obtaining the path
$$
P_3 = [S' \rightarrow \cdot S], \ldots, [C_t \rightarrow \cdot C_{t-1}\gamma_t], \ldots,
     [C_2 \rightarrow C_1\cdot\gamma_2]. 
$$
This kind of derivations ist continued obtaining the path
$$
P_{t-1} = [S' \rightarrow \cdot S], \ldots, [C_t \rightarrow C_{t-1}\cdot\gamma_t]. 
$$
Theorem 1 implies that all these expansions and reductions are performed with respect to all 
paths which are suitable for the end node $[A \rightarrow \cdot]$ in ${\cal G}'$. 
Hence, it is easy to prove by induction that each path $P$ from the start node to 
$[A \rightarrow \cdot]$ with $a \in FIRST_1(suf(P))$ in ${\cal G}'$ can be written as 
$P = P'Q$, where
$$
Q =  [C_{t-1} \rightarrow \cdot C_{t-2}\gamma_{t-1}], \ldots, [C_2 \rightarrow \cdot C_1\gamma_2], 
     [C_1 \rightarrow \cdot A\gamma_1], [A \rightarrow \cdot].
$$
After performing the expansions and reductions with respect to the path $Q$ in
the graph ${\cal G}'$ we obtain a graph ${\cal G}_1$ which contains for each such a path 
$P = P'Q$ the path $P''$, where $P''$ is obtained from $P'$ by moving the dot in the last node 
of $P'$ one position to the right. The path $P_{t-1}$ is obtained from the path $P_0$. Note that
$[C_t \rightarrow C_{t-1}\cdot\gamma_t]$ is an end node of ${\cal G}_1$.
Now, an expansion step is performed. The expansion step transforms the graph ${\cal G}_1$ to
a graph ${\cal G}'_1$. Two subcases are possible:

\medskip
\noindent
{\em Case 2.1\/}: There is an end node $[A' \rightarrow \cdot]$ with ${\cal G}'_1$
                  contains a path which is suitable for $[A' \rightarrow \cdot]$.

\smallskip
Theorem 1 implies that $[A' \rightarrow \cdot]$ is the unique end node for which a 
suitable path in ${\cal G}'_1$ exists. The same consideration as above shows that there is a path 
$Q'$ such that each path $P''$ in ${\cal G}'_1$ which is suitable for 
$[A' \rightarrow \cdot]$ can be written as $P'' = RQ'$. Note that for
different such paths $P''$, the paths $R$ can be different. After performing the expansions and 
reductions with respect to the path $Q'$ in the graph ${\cal G}'_1$ we obtain a graph ${\cal G}_2$,
and so on.

\medskip
\noindent
{\em Case 2.2\/}: There is no end node $[A' \rightarrow \cdot]$ with ${\cal G}'_1$
                  contains a path which is suitable for $[A' \rightarrow \cdot]$.

\smallskip
Then there is at least one readable end node and the readable end nodes are exactly the end nodes
for which the graph ${\cal G}'$ contains a suitable path. Then we are in Case 1.

\medskip
Next we will analyze the used time and space of the simulation of $LR(k)$-$M_G$. Let $ld(input)$
denote the length of the derivation of the input. The insertion and the deletion of all nodes
and edges which correspond to reductions performed during the simulation of $LR(k)$-$M_G$ can
be counted against the corresponding production in the derivation of the input. We have shown
that with respect to all paths from the start node to an end node in the current graph, the same
reductions have been performed. Hence, the total time used for such nodes and edges is 
$O(ld(input))$. Besides these nodes and edges, $O(|G|)$ nodes and edges are inserted during a phase. 
Note that during a phase, such nodes and edges are inserted at most once. Hence, the total time used 
for all expansions which do not correspond to reductions is bounded by $O(n|G|)$. 
During a reversed topological search, the time used for the visit of a node or an edge is zero or 
constant. If during the search a node is visited, the node takes part on an expansion, is deleted 
or its dot is moved one position to the right. Hence, the total time used for nodes and edges
during a reversed topological search which do not correspond to a reduction is bounded by 
$O(n|G|)$.

\smallskip
During the reversed topological searchs we have to decide if $\varepsilon$ or the next unread
input symbol $a$ is contained in $FIRST_1(A_j)$ where $A_j \in V$. This is trivial for 
$A_j \in \Sigma$. For $A_j \in N$ we need a representation of $FIRST_1(A_j)$. A possible 
representation is an array of size $|\Sigma| + 1$ such that each decision can be made in constant 
time. Then, we need for each variable in $N$ an additional space of size $O(|\Sigma|)$.
Altogether, we have proved the following theorem:
\begin{theo}
Let $G = (V,\Sigma,P,S)$ be an $LR(1)$-grammar. Let $ld(input)$ denote the length of the 
derivation of the input. Then there is an extended $LR(1)$-parser $P_G$ for $G$ which has 
the following properties:
\vspace{-0.2cm}
\begin{itemize}
\item[i)]
The size of the parser is $O(|G| + |N||\Sigma|)$.
\item[ii)]
$P_G$ needs only the additional space for the manipulation of a directed graph of size $O(|G|n)$
where $n$ is the length of the input.
\item[iii)]
The parsing time ist bounded by $O(ld(input) + n|G|)$. 
\end{itemize}
\end{theo}
Next, we will extend the solution developed for $k = 1$ to larger $k$. A lookahead 
$u := x_1x_2 \ldots x_k$ of length $k$ has $k$ proper prefixes $\varepsilon, x_1, x_1x_2, 
\ldots, x_1x_2\ldots x_{k-1}$. Let $P$ be a path from the start 
node of $\cal{G}$ to an end node $[A \rightarrow \cdot]$ or $[A \rightarrow 
\alpha\cdot a\beta]$ and let $suf(P) = A_1A_2 \ldots A_r$. Note that 
$u \in FIRST_k(A_1A_2 \ldots A_r)$ iff for all $1 < i \leq r$ there is $u \in FIRST_k(A_1A_2
\ldots A_{i-1})$ or there is a proper prefix $u'$ of $u$ such that $u' \in FIRST_k(A_1A_2 \ldots 
A_{i-1})$ and $u'' \in FIRST_k(A_iA_{i+1} \ldots A_r)$ where $u = u'u''$. For the decision if 
$u \in FIRST_k(suf(P))$ we consider $A_1A_2 \ldots A_r$ from left to right. Assume that $A_j$ is
the current considered symbol. If $j = 1$ then we have to compute all prefixes of $u$ which are
contained in $FIRST_k(A_1)$. If no such a prefix exists then $u \not\in FIRST_k(A_1A_2 \ldots 
A_r)$. Assume that $j > 1$. Let $U := \{u'_1,u'_2, \ldots, u'_s\} \not= \emptyset$ be the set 
of proper prefixes of $u$ which are contained in $FIRST_k(A_1A_2 \ldots A_{j-1})$. For $1 \leq i 
\leq s$ let $u = u'_iu''_i$. We have to compute all prefixes of $u''_i$ which are contained in 
$FIRST_k(A_j)$. If no prefix of $u''_i$ is contained in $FIRST_k(A_j)$ then
$u'_i$ cannot be extended to the lookahead $u$ with respect to $FIRST_k(A_1A_2 \ldots A_r)$ and
needs no further consideration. Then the prefixes of $u$ contained in 
$FIRST_k(A_1A_2 \ldots A_j)$ are obtained by the concatenation of $u'_i$ and all prefixes of 
$u''_i$ in $FIRST_k(A_j)$ for $1 \leq i \leq s$.

\smallskip
For the decision if there is an end node of $\cal{G}$ such that there is a path $P$ from the
start node to this end node with $u \in FIRST_k(suf(P))$, a reversed topological search on 
${\cal G}'$ is performed. Since the length of the lookahead is larger than one, a graph search 
has also to be performed with respect to readable end nodes. Furthermore, we have to
transport from a node $v$ to its predecessors a list of proper prefixes of $u$ where each
prefix is associated with a unique reducible or readable end node. For a node $v$ we denote its 
list of prefixes of $u$ by $L(v)$. A proper prefix $u'$ of $u$
associated with an end node $w$ is contained in the list iff there is a path $Q$ from $v$ to 
$w$ such that $u' \in FIRST_k(suf(Q))$. 
Only nodes which correspond to variables have outdegree larger than one. For such nodes we obtain 
one list by the union of the lists corresponding to its direct successors. Since $LR(k)$-grammars 
are unambiguous, the same proper prefix of $u$ is not contained in two different lists with respect 
to the direct successors of a node.
Assume that we enter the node $C$ over an outgoing edge $e$ and there is a closing edge 
$([B \rightarrow \cdot C\gamma],C)$.
Before the continuation of the reversed topological search at the node $C$, the list $L(C)$ 
has to be transported to the node $[B \rightarrow \cdot C\gamma]$ and the search is 
continued at the node $[B \rightarrow \cdot C\gamma]$. Note that this node is a successor of the 
node $C$ in the graph ${\cal G}'$. But it cannot happen that we visit the node $C$ again with a
prefix $u'$ of $u$ which is already contained in the list $L(C)$.
Otherwise, the grammar would be ambiguous. Hence, the node $C$ can only be reentered with prefixes
of $u$ which are not already in the list. During the last run through the cycle,
either one finds a path $P$ such that $u \in FIRST_k(suf(P))$ or one detects that no such a path 
using the cycle again exists before reaching the node $C$. Hence, the number of continuations
of the reversed topological search at the node $[B \rightarrow \cdot C\gamma]$ is bounded by
$k-1$. This observation implies that, although the graph ${\cal G}'$ may contain some
cycles, for the reversed topological search, the graph can be considered as an acyclic graph.
After the treatment of all outgoing edges of the node $C$ and all closing edges with end node $C$, 
the final list $L(C)$ is computed. Then, the reversed topological search is 
continued at the node $C$.

\smallskip
Assume that during a reversed topological search, the symbol $A_j \in V$ is considered and that
$U = \{u'_1,u'_2, \ldots, u'_s\}$ is the set of proper prefixes of the current lookahead $u$
which belongs to this consideration. This means that for all $1 \leq i \leq s$ there is a path $Q_i$ 
from the point under consideration in the graph to an end node such that $u'_i \in FIRST_k(suf(Q_i))$. 
For $1 \leq i \leq s$ let $u = u'_iu''_i$. We have to compute all prefixes of $u$ which are 
contained in $FIRST_k(suf(Q_i)A_j)$ for any $1 \leq i \leq s$. Let $U'$ denote the set of these
prefixes. A prefix $\bar{u}$ of $u$ is contained in $U'$ iff there exists $1 \leq i \leq s$ such 
that
\begin{enumerate}
\item
$\bar{u} = u'_i\bar{u}_i$ and $\bar{u}_i \in FIRST_k(A_j)$, or
\item
$\bar{u} = u'_i\bar{u}_i = u$ and $\bar{u}_i$ is prefix of an element in $FIRST_k(A_j)$.
\end{enumerate}
Since $LR(k)$-grammars are unambigious, for all $\bar{u} \in U'$ there is exactly one
$i \in \{1,2, \ldots,s\}$ such that $\bar{u} \in FIRST_k(suf(Q_i)A_j)$.
Note that all information needed for the computation of $U'$ depends only on the lookahead $u$, 
the set $U$ and $FIRST_k(A_j)$. Hence, in dependence on all possible $u$, $U$ and $A_j$, we can 
precompute the corresponding sets $U'$.  

\smallskip
Instead of storing the elements of $U$ we can store the lengths of these prefixes. This can be 
realized by a binary vector of length $k$. The $i$-th component of the vector is one iff the 
prefix of length $i - 1$ is in $U$.
Let $v(U)$ denote the vector corresponding to the list $U$ of prefixes of $u$.
 With respect to each component of the vector with value one, we have to specify the unique
end node of the graph associated with the corresponding proper prefix of $u$. If we number the
posssible end nodes then at most $O(\log |G|)$ bits are needed for each spezification.

We can represent the set of vectors corresponding to all possible $U'$ by a table where in 
dependence of the current lookahead $u$, the current $v(U)$ and the symbol $A_j \in V$
under consideration we get the vector $v(U')$. For each component $i$ in $v(U')$ which has the 
value one we need the spezification of the component $l$ such that $x_1x_2 \ldots x_l \in U$ and
$x_{l+1}x_{l+2} \ldots x_i \in FIRST_k(A_j)$ or $x_{l+1}x_{l+2} \ldots x_i$ is a prefix of an
element in $FIRST_k(A_j)$ in the case $i = k$. Then, the end node corresponding to $x_1x_2 \ldots
x_l$ in $U$ is the end node which corresponds to $x_1x_2 \ldots x_i$ in $U'$.
Let $\#LA$ denote the number of the relevant lookaheads with respect to the $LR(k)$-grammar $G$.
Obviously, $\#LA \leq |\Sigma|^k$. Then, the size of the table above is $O(\#LA |V| 2^k k \log k)$.
If $A_j \in \Sigma$ then $U'$ can easily be computed from $U$ in $O(k)$ time. 
This would reduce the size of the table to $O(\#LA |N| 2^k k \log k)$.
During the reversed topological search, the time used for the visit of a node or an edge is $O(k)$.
Hence, the used time increases to $O(ld(input) + k|G|n)$.
Altogether, we have proved the following theorem:
\begin{theo}
Let $G = (V,\Sigma,P,S)$ be an $LR(k)$-grammar. Let $\#LA$ denote the number of possible lookaheads
of length $k$ with respect to $G$ and let $ld(input)$ denote the length of the derivation of the
input. Then there is an extended $LR(k)$-parser $P_G$ for $G$ which has the following properties:
\vspace{-0.2cm}
\begin{enumerate}
\item
The size of the parser is $O(|G| + \#LA |N| 2^k k \log k)$.
\item
$P_G$ needs only the additional space for the manipulation of a directed graph of size $O(|G|n)$
where $n$ is the length of the input.
\item
The parsing time ist bounded by $O(ld(input) + k|G|n)$.
\end{enumerate}
\end{theo}

\subsection{Lookaheads of large size}

If the size $k$ of the lookahead is large then the size $\#LA |N| 2^k k \log k$ of the table
described in Section 7.1 can be too large. Hence, we describe an implementation of
$LR(k)$-$M_G$ without the precomputation of these tables. For getting an efficient implementation,
the parser $P_G$ contains for each variable $X \in N$ the trie $T_k(X)$ which represents the set
$FIRST_k(X)$. Assume that during a reversed topological search, the symbol $A_j \in V$ is considered 
and that $U = \{u'_1,u'_2, \ldots, u'_s\}$ is the set of proper prefixes of the current lookahead 
$u := x_1x_2 \ldots x_k$
which belongs to this consideration. This means that for all $1 \leq i \leq s$ there is a path $Q_i$ 
from the point under consideration in the graph to an end node such that $u'_i \in FIRST_k(suf(Q_i))$. 
For $1 \leq i \leq s$ let $u = u'_iu''_i$. We have to compute the set $U'$ of all prefixes of $u$ 
which are contained in $FIRST_k(suf(Q_i)A_j)$ for any $1 \leq i \leq s$. Instead of using a 
precomputed table, the parser $P_G$ considers iteratively all lengths $|u_1'|, |u_2'|, \ldots,|u_s'|$.
Let $q$ be the current considered length. Two cases can arise:

\smallskip
\noindent
{\em Case 1:\/} $A_j \in \Sigma$. 

\smallskip
If $A_j = x_{q+1}$ then $FIRST_k(A_1A_2 \ldots A_j)$ contains the prefix $x_1 \ldots x_{q+1}$ 
of the lookahead $u$. Hence, we increase the current considered length $q$
by one. Otherwise, $P_G$ is in a dead end with respect to the prefix of length $q$ of $u$ 
such that the length $q$ can be deleted with respect to the path $P$.

\smallskip
\noindent
{\em Case 2:\/} $A_j \in N$. 

\smallskip
Our goal is to determine all prefixes of $u(q) := x_{q+1}x_{q+2} \ldots x_k$ which can be derived 
exactly from $A_j$. This means that we have to compute all prefixes $u''$ of $u(q)$ which are also 
prefix of an element in $FIRST_k(A_j)$. For doing this, $P_G$ starts to read $x_{q+1}x_{q+2} 
\ldots x_k$ and, simultaneously, to follow the corresponding path in $T(A_j)$, starting at the root,
until the maximal prefix $\tilde{u}$ of $u(q)$ in $T_k(A_j)$ is determined. If $|\tilde{u}| = k-q$
then $u \in FIRST_k(A_1A_2 \ldots A_j)$ and $P_G$ knows that the corresponding $LR(k)$-item is valid
for a path $P$. If $|\tilde{u}| < k-q$ then we have to derive from $A_j$ a prefix $u''$ of 
$\tilde{u}$. Hence, it is useful if $P_G$ has direct access to all such prefixes. For getting this, 
every node $v \in T_k(X)$, $ X \in N$ contains a pointer to the node $w \in T_k(X)$ such that
\begin{itemize}
\item[a)] 
$s(w) \in FIRST_k(X)$, and
\item[b)]
$w$ is the last node $\not= v$ on the path from the root to $v$ which fulfills a).
\end{itemize}
For each such a prefix $u''$, $P_G$ stores $q + |u''|$. Note that $P_G$ already 
read $l-q$ of these $k-q$ symbols where $l$ is the length of the prefix of the 
lookahead already read. We do not want to read these symbols of the input again.
Hence, $P_G$ needs the possibility of direct access to the ``correct'' node in $T_k(A_j)$ with 
respect to the read prefix of the next $k-q$ symbols. 
For getting this direct access, we extend $P_G$ by a trie $T_G$ representing the set
$\Sigma^k$. Moreover, $P_G$ manipulates a pointer $P(T_G)$ which always points to the node
$r$ in $T_G$ with $s(r)$ is the prefix of the lookahead $u$ already read.
For $v \in T_G$ let $d(v)$ denote the depth of $v$ in $T_G$ and $s_i(v)$, 
$0 \leq i < d(v)$ denote the suffix of $s(v)$ which starts with the $(i+1)$st
symbol of $s(v)$. Every node $v \in T_G$ contains for all $A \in N$ and
$1 \leq i < d(v)$ a pointer $P_{i,A}(v)$ which points to the node $w \in T_k(A)$
such that $s(w)$ is the maximal prefix of an element of $FIRST_k(A)$ which is
also a prefix of $s_i(v)$. Using the pointer $P_{q,A_j}(v)$, where $v$ is the node to which 
$P(T_G)$ points, $P_G$ has direct access to the correct node $w$ in $T_k(A_j)$. 

\smallskip
If $s(w) \not= s_q(v)$ then $s(w)$ is the maximal prefix of $s_q(v)$ which is 
prefix of an element of $FIRST_k(A_j)$. 
For every prefix $u''$ of $s(w)$ with $u'' \in FIRST_k(A_j)$, the parser $P_G$ knows
that $x_1x_2 \ldots x_qx_{q+1}x_{q+2} \ldots x_{q+|u''|} \in FIRST_k(A_1A_2 \ldots A_j)$.
Hence, $P_G$ stores the length $q + |u''|$. If no such $u''$ exists, then $P_G$ is in a
dead end with respect to the length $q$ such that the length $q$ can be deleted from the list.

\smallskip
If $s(w) = s_q(v)$ then $P_G$ continues to read the rest of the lookahead $u$
and, simultaneously, follows the corresponding path in $T_k(A_j)$, starting at 
the node $w$. If the whole lookahead $u$ is read and $u \in FIRST_k(A_1A_2 \ldots 
A_j)$ then the process is terminated and $P_G$ knows that the $LR(k)$-item 
$[A \rightarrow \cdot, u]$ is valid for a path $P$. Otherwise, 
if a proper prefix $u''$ of $u$ is in $FIRST_k(A_1A_2 \ldots A_j)$ then 
$P_G$ continues the reversed topological search.

\smallskip
Next we want to bound the size of $P_G$. By construction, $P_G$ contains
$|N|+1$ tries, the trie $T_G$ and the tries $T_k(A)$, $A \in N$. Each trie consists of at 
most $2|\Sigma|^k$ nodes. Since we only need nodes which correspond to possible lookaheads,
this bound reduces to $k\#LA$. Every node in a trie $T_k(A)$, $A \in N$ contains one pointer.
Each node in $T_G$ contains for every $A \in N$ at most $k$ pointers which points to a node
in $T_k(A)$. Therefore, the total number of pointers in $T_G$ is bounded by 
$\min \{\#LA|N|k^2, |\Sigma|^k|N|2k\}$. Hence, all tries need $O(\min\{\#LA|N|k^2, |\Sigma|^k|N|k\})$ 
space. 
Altogether, we have proved the following theorem:
\begin{theo}
Let $G = (V,\Sigma,P,S)$ be an $LR(k)$-grammar. Let $\#LA$ denote the number of relevant lookaheads
of length $k$ with respect to $G$ and let $ld(input)$ denote the length of the derivation of the
input. Then there is an extended $LR(k)$-parser $P_G$ for $G$ which has the following properties:
\vspace{-0.2cm}
\begin{enumerate}
\item
The size of the parser is $O(|G| + \min\{\#LA|N|k^2, |\Sigma|^k|N|k\})$.
\item
$P_G$ needs only the additional space for the manipulation of a directed graph of size $O(|G|n)$
where $n$ is the length of the input.
\item
The parsing time ist bounded by $O(ld(input) + k|G|n)$.
\end{enumerate}
\end{theo}

\section{Experimental Results}

In his diploma thesis, Heinz-Christian Steinhausen \cite{St} has implemented a parser
generator for extended $LR(k)$-parser. For the comparision of the generated parsers
with canonical parsers of the same grammars he used the parser generators Bison \cite{Bison}
and MSTA \cite{Maka}. Bison generates for a given $LALR(1)$-grammar a canonical $LALR(1)$-parser.
MSTA can be used for $LALR(k)$- and for $LR(k)$-grammars.

The canonical $LR(k)$-parser contains precomputed parsing tables such that the next step of the parser
can be uniquely determined by table lookups. As data structure during the parsing only a stack is
needed. What we pay for the precomputation of the whole information
such that the unique next move of the parser can be decided only by the use of the parsing table is
the size of the table which can be exponential in the size of the underlying grammar. Hence, for
using of canonical $LR(k)$-parser one has to take care that the structur of the grammar allows to
get a parser with a parsing table of moderate size. 

In contrast to canonical parsers, the extended $LR(k)$-parser computes some portion of the needed
information for getting the unique derivation during the parsing of the program. For doing this
more complicated data structures than a stack are needed. Instead of a stack we manipulate a directed 
graph of size $O(|G|n)$ where $n$ is the length of the input. What we gain is that the size of the parser
is always polynomial in the size of the underlying grammar. What we pay is the increased time used for 
parsing the program since the manipulation of a graph is more time consuming than simple table lookups.

Hence, it is no surprise that for grammars which are optimized such that the canonical parser works
well, the size of the extended parser is larger than the size of the canonical parser and the parsing 
time of the extended parser is much larger than the parsing time of the canonical parser. 
Hence, the techniques of the extended parser should only be used if the canonical parser does not work
well. To get an impression of the usefullness of the extended $LR(k)$-parser, Steinhausen has considered
the family $G_n = (V_n,\Sigma_n,P_n,S)$, $n \in \mathbb{N}$ of $LR(0)$-grammars where
$$\begin{array}{lll}
P_n: & S \rightarrow A_i                  & (1 \leq i \leq n), \\ 
     & A_i \rightarrow a_jA_i             & (1 \leq i \not= j \leq n), \\
     & A_i \rightarrow a_iB_i \mid b_i    & (1 \leq i \leq n), \\
     & B_i \rightarrow a_jB_i \mid b_i    & (1 \leq i,j \leq n).
\end{array}$$
The size $|G_n| = m$ equals $6n^2 + 5n$.
Earley \cite{Ea} has established the family of grammars $G_n$ as an example where the size of the
canonical $LR(0)$-parser grows exponentially in $n$. Ukkonen \cite{Uk} has proved that the size
of every $LR(k)$-, $SLR(k)$- or $LALR(k)$-parser for $G_n$ is larger than $2^{c\sqrt{m}}$ for some constant
$c > 0$. 

Table \ref{t:1} compares the sizes of the generated parsers for the grammars $G_n, 1 \leq n \leq 10$.
With respect to $G_{20}$, the extended $LR(0)$-parser of size 714 kB has been generated within three seconds.
After 24 hours, Bison has not computed a canonical parser. After seven minutes, MSTA has terminated the computation
because of storage overflow. The three parsers generated for the grammar $G_{10}$ have been testet with the input
$x = a_2^{9998}a_1b_1$ which has length 10000. The extended $LR(0)$-parser terminates after 371 ms. Both canonical
parsers terminates after 2 ms with an error message. The error message of the parser generated by Bison was
``memory exhausted'' and the error message of the parser generated by MSTA was ``states stack is overfull''.

\begin{table}
\centering
  {\begin{tabular}{l|r|r|r}
  & ExtendedLR & Bison & MSTA \\\hline
$G_1$ & 143 & 39 & 15 \\
$G_2$ & 149 & 41 & 19 \\
$G_3$ & 157 & 45 & 27 \\
$G_4$ & 167 & 54 & 45 \\
$G_5$ & 181 & 76 & 92 \\
$G_6$ & 197 & 127 & 214 \\
$G_7$ & 216 & 255 & 540 \\
$G_8$ & 237 & 568 & 1395 \\
$G_9$ & 261 & 1343 & 3559 \\
$G_{10}$ & 288 & 3199 & 8760 \\
$G_{20}$ & 714 &    no   &  no     \\
\end{tabular}}
\caption{\label{t:1}Parser sizes in kB.}
\end{table}

\section{Conclusion}

We have constructed for $LR(k)$-grammars extended $LR(k)$-parsers of polynomial size. Hence, for
small sizes of the lookahead (e.g. $k < 10$), $LR(k)$-grammars can be used for the definition of
the syntax of programming languages. Moreover, if the number of relevant lookaheads is not
too large, $LR(k)$-grammars can also be used for larger sizes of the lookahead. 
These possibilities open some new research directions, for instance:
\vspace{-0.2cm}
\begin{enumerate}
\item
The construction of parser generators for extended $LR(k)$-parsers.
\item
The combination of canonical and extended $LR(k)$-parsers such that the advantages of both
concepts are maintained and the disadvantages are omitted.
\item
The development of new concepts for programming languages which use $LR(k)$-grammars for larger 
$k$.
\item
The extension of the method to natural languages (see \cite{To}).
\end{enumerate}

\section*{Acknowledgments}

I thank Heinz-Christian Steinhausen for the implementation of the first parser generator for
extended $LR(k)$-parsers and his experiments to compare parsers generated by his
parser generator with parsers generated by Bison and by MSTA.


\end{document}